\documentclass[prl,twocolumn,amsmath,amssymb,superscriptaddress, showpacs]{revtex4-1}

\usepackage{amsmath} 
\usepackage{graphicx} 
\usepackage{verbatim} 
\usepackage{color}     
\usepackage[usenames,dvipsnames,svgnames,table]{xcolor}
\usepackage{subfigure}  
\usepackage[hidelinks]{hyperref} 
\usepackage{multirow}
\usepackage{float}
\usepackage{braket}
\usepackage{siunitx}
\usepackage{bbm}
\usepackage{microtype}  
\usepackage{bibunits}

\defaultbibliography{pqgbib}

\graphicspath{{images/}}
\newcommand{\ind}[1]{\ensuremath{_{\text{#1}}}}
\newcommand{\dni}[1]{\ensuremath{^{\text{#1}}}}

\newcommand{\unit}[1]{\ensuremath{\,\mathrm{#1}}}

\newcommand{\ZZ}{\textsl{ZZ}}
\newcommand{\CZ}{\textsl{CZ}}

\newcommand{\puttitle}{Implementation of Conditional-Phase Gates based on tunable ZZ-Interactions}

\begin{document}

\title{\puttitle}
\author{Michele C. Collodo}
\thanks{These authors contributed equally\\michele.collodo@phys.ethz.ch}
\author{Johannes Herrmann}
\thanks{These authors contributed equally\\michele.collodo@phys.ethz.ch}
\author{Nathan Lacroix}
\author{Christian Kraglund Andersen}
\author{Ants~Remm}
\author{Stefania Lazar}
\author{Jean-Claude Besse}
\author{Theo Walter}
\author{Andreas Wallraff}
\author{Christopher Eichler}
\affiliation{Department of Physics, ETH Zurich, CH-8093 Zurich, Switzerland}

\date{\today}


\begin{abstract}
\noindent High fidelity two-qubit gates exhibiting low crosstalk are essential building blocks for gate-based quantum information processing. In superconducting circuits two-qubit gates are typically based either
on RF-controlled interactions or on the in-situ tunability of qubit frequencies. Here, we present an alternative
approach using a tunable cross-Kerr-type \ZZ -interaction between two qubits, which we realize by a flux-tunable coupler element.
We control the \ZZ -coupling rate over
three orders of magnitude to perform a rapid ($38 \unit{ns}$), high-contrast, low leakage ($0.14 \, \%$) conditional-phase \CZ \ gate with a fidelity of $97.9\,\%$ without relying on the resonant interaction with a non-computational state. Furthermore, by exploiting the direct nature of the \ZZ -coupling, we easily access the entire conditional-phase gate family by adjusting only a single control parameter. 
\end{abstract}

\maketitle

\begin{bibunit}[apsrev4-1]
Superconducting circuits have become one of the most advanced physical systems for building quantum information processing devices and for performing high-fidelity operations for control and measurement~\cite{kjaergaard_superconducting_2020, arute_quantum_2019, kandala_hardware-efficient_2017, rosenblum_fault-tolerant_2018, dicarlo_demonstration_2009, andersen_repeated_2019}. 
While single qubit gates are routinely realized with very high fidelity~\cite{motzoi_simple_2009}, achieving similar performance in two-qubit gates remains an outstanding challenge.
Multiple criteria have been established to assess the quality of two-qubit gate schemes, which include the gate error~\cite{corcoles_process_2013}, the susceptibility to leakage out of the computational subspace~\cite{rol_fast_2019}, the duration of gates~\cite{barends_diabatic_2019}, the residual coupling during idle times~\cite{chen_qubit_2014,krinner_benchmarking_2020,mundada_suppression_2019, ku_suppression_2020}, as well as the flexibility of realizing different types of two-qubit gates with continuously adjustable gate parameters~\cite{ganzhorn_gate-efficient_2019, foxen_demonstrating_2020, abrams_implementation_2019, lacroix_improving_2020}.

In general, the realization of two-qubit gates relies on a controllable coupling mechanism, which in the case of superconducting qubits is usually a transversal coupling of the form $\sigma\ind{x}\dni{(1)}\sigma\ind{x}\dni{(2)}$, where $\sigma\ind{x}\dni{(i)}$ are Pauli-operators along the $x$-axis transversal to the qubit quantization axis $z$. 
Common methods to dynamically control this transverse coupling use fast dc flux pulses to either bring qubit states into and out of resonance~\cite{dicarlo_demonstration_2009, barends_superconducting_2014, dewes_characterization_2012} or to tune the effective coupling rate directly using a tunable coupler element~\cite{chen_qubit_2014, foxen_demonstrating_2020, yan_tunable_2018, li_tunable_2020, barends_diabatic_2019}.
Alternative methods are based on driving sideband transitions, induced either by parametric flux-modulation ~\cite{mckay_universal_2016, ganzhorn_gate-efficient_2019, mundada_suppression_2019, reagor_demonstration_2018, caldwell_parametrically_2018, chu_realization_2019, hong_demonstration_2020, abrams_implementation_2019} or by microwave charge drives~\cite{chow_simple_2011, chow_implementing_2014, sheldon_procedure_2016}.

Steady improvements of gate fidelities have recently began to reveal inherent challenges of gates based on transverse coupling:
While SWAP-type gates are implemented through the direct coupling of computational states, Conditional-\textsl{Z} (\CZ) gates exploit the coupling to an auxiliary, non-computational state, making it prone to leakage errors.
Furthermore, residual couplings during idle times may be a source of
correlated errors, which are especially detrimental on larger devices. This effect can be suppressed by increasing the gate contrast, defined as the ratio between the interaction rates during the gate and during idle times. In an effort to overcome both leakage errors and idle coupling, net-zero pulse parametrization schemes~\cite{rol_fast_2019} and high-contrast coupling mechanisms~\cite{mundada_suppression_2019, ku_suppression_2020, noguchi_fast_2020} have been investigated more recently.

Here, we address both aforementioned challenges by implementing controlled phase gates based on an in-situ tunable \ZZ -interaction described by the Hamiltonian
\begin{align*}
    \mathcal{H}\ind{eff}/\hbar &= \frac{1}{2}\sum_{i=1,2} 
     \left(\omega_i+\frac{\alpha\ind{\ZZ}}{2}\right) \, {\sigma^{(i)}\ind{z}} 
    + \,\frac{\alpha\ind{\ZZ}}{4} \, \sigma\dni{(1)}\ind{z} \sigma\dni{(2)}\ind{z}.
\end{align*}
Here, $\omega\ind{1,2}$ are the qubit frequencies, $\alpha\ind{\ZZ}$ is the tunable cross-Kerr \ZZ -interaction rate, and $\sigma\dni{(1),(2)}\ind{z}$ are the Pauli-operators.

Our approach ensures direct control of the acquired conditional phase, without relying on excitation transfer or sideband transitions, and thus features an inherent resilience to leakage and crosstalk. Furthermore, it allows to freely choose a target conditional phase, without having to recalibrate gate parameters for population recovery.
Moreover, the \ZZ -interaction is only weakly dependent on the qubit detuning, allowing for a flexible choice of frequency configurations to avoid frequency crowding, which is of particular relevance when scaling up the number of qubits.

\begin{figure}[t!] 
\centering
\includegraphics[width = 0.48\textwidth]{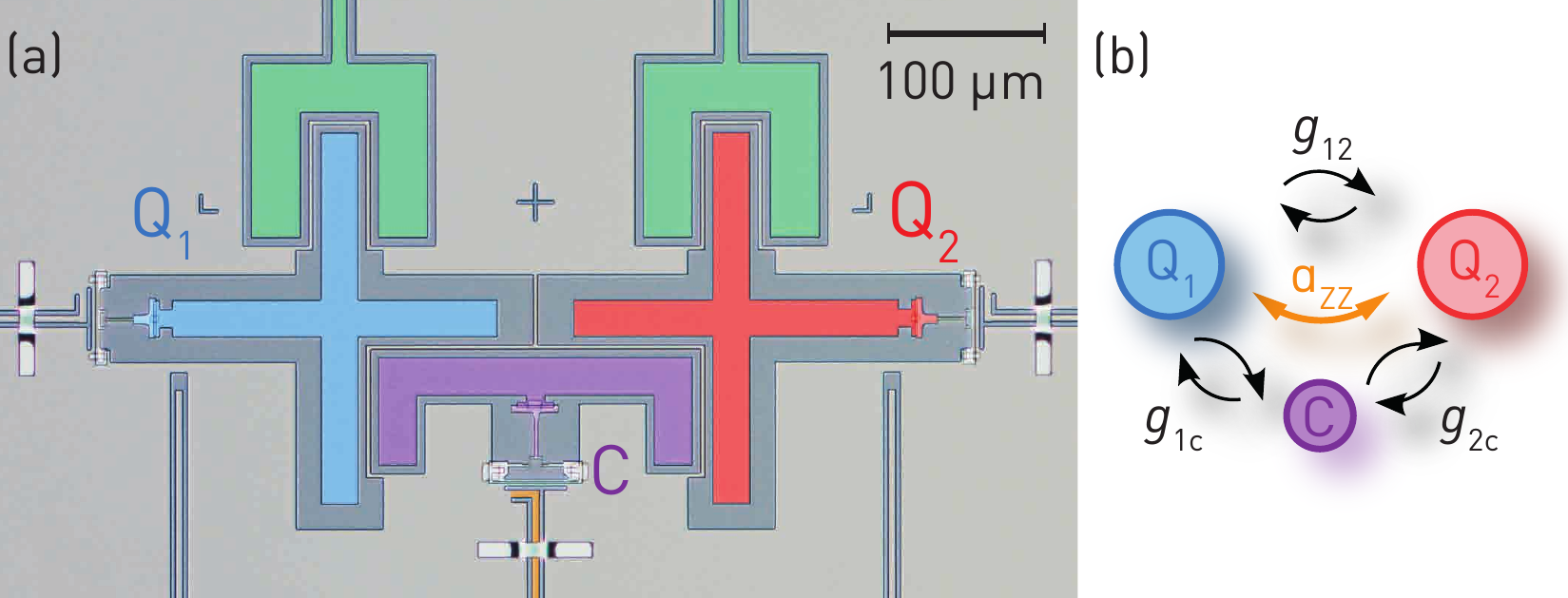}
\caption{
(a) False-colored micrograph of the sample, featuring two transmons used as computational qubits (Q$_1$, blue; Q$_2$, red), a coupler transmon (C, purple), its flux line for frequency control (orange), the readout circuitry (green) and additional charge- and flux control lines of Q$_1$, Q$_2$.
(b) Schematic illustrating the transversal interactions $g\ind{12}$, $g\ind{1c}$, $g\ind{2c}$ between the computational qubits and the coupler qubit. The \ZZ -coupling (orange arrow) arises from the hybridization of all three nonlinear modes and can be tuned as a function of the coupler frequency.
}
\label{fig1}
\end{figure}

We implement this coupling mechanism in the superconducting circuit shown in Fig.~\ref{fig1}. It is composed of two superconducting Xmon-style single island transmons (blue, red) operating as computational qubits (Q$_1$,Q$_2$) and a single island transmon (purple) operating as a nonlinear coupling element (C). All three elements are frequency tunable via an external magnetic flux, which we apply through dedicated on-chip flux control lines (orange). We park the qubits at $(\omega_1, \omega_2, \omega\ind{c}) / 2\pi = (5.038, 5.400, 7.612) \unit{GHz}$ during idle operation.
The coupling between Q$_1$ and Q$_2$ is mediated by a direct capacitance yielding a simulated coupling rate $g\ind{12} / 2 \pi = 33 \unit{MHz}$ as well as a second order capacitive path with rates $(g\ind{1c}, g\ind{2c}) / 2 \pi = (265, 274) \unit{MHz}$ comprising the coupler.
The \ZZ-interaction arises from the interplay between these two coupling channels and the resulting hybridization~\cite{strauch_multiphoton_2018} of the three participating local modes with measured anharmonicities $(\alpha\ind{1}, \alpha\ind{2}, \alpha\ind{c}) / 2 \pi = (-240, -238, -269) \unit{MHz}$. For the idle configuration we calculate an effective residual transverse coupling between Q$_1$, Q$_2$ of $J / 2 \pi \approx -2 \unit{MHz} \ll (\omega_2 - \omega_1) / 2 \pi$ (see Supplemental Material~\cite{supplemental} for a detailed description of the circuit quantization).

We use the frequency of the coupler element as a control parameter to tune $\alpha\ind{\ZZ} (\omega\ind{c})$. This emphasizes the role of the coupler as an external control system and requires it to remain in the ground state at all times.
Our scheme is compatible with weakly tunable or fixed frequency qubits, as solely the interfacing coupler element requires frequency tunability, mitigating the influence of flux noise induced dephasing~\cite{hutchings_tunable_2017}.

\begin{figure}[t!] 
\centering
\includegraphics[width = 0.48\textwidth]{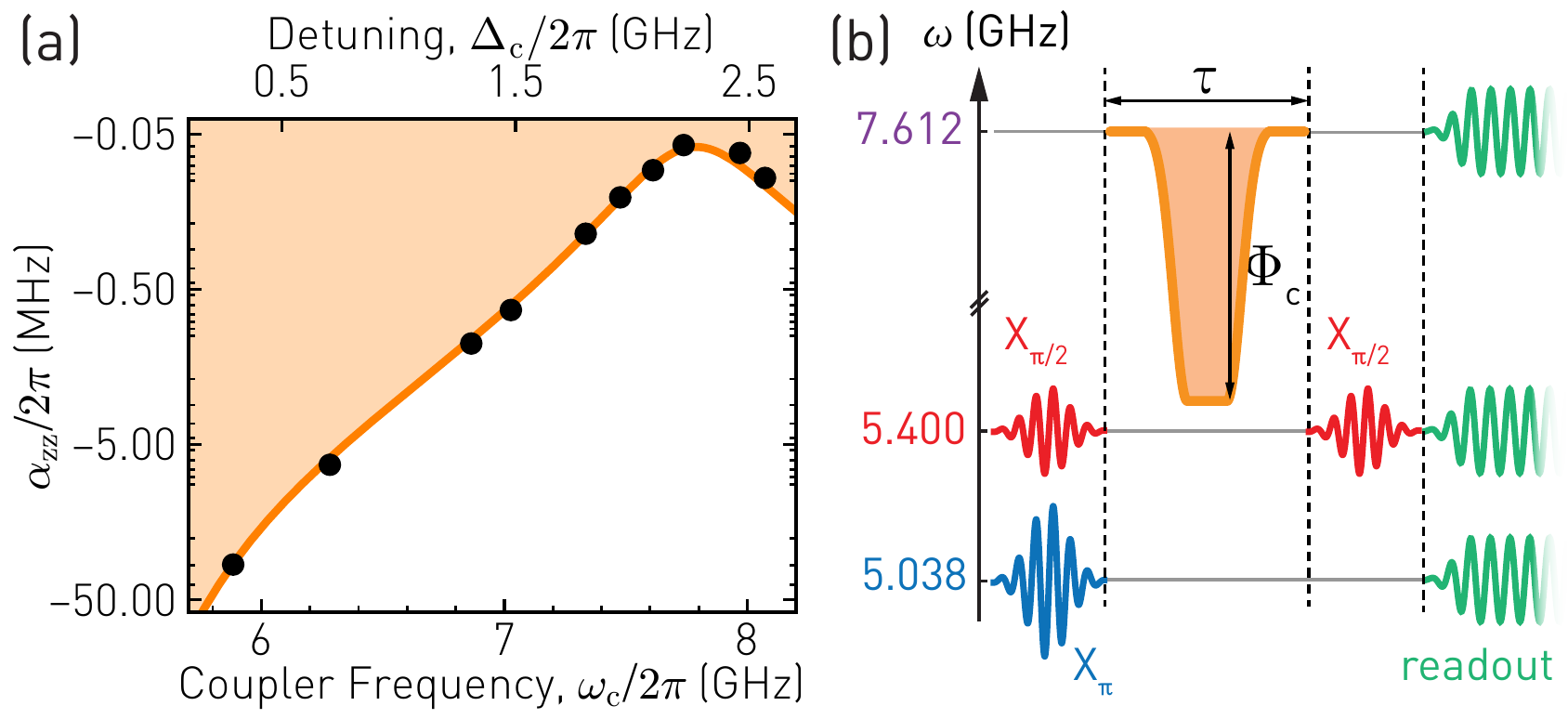}
\caption{
(a) Measured \ZZ-coupling strength $\alpha\ind{ZZ}$ \textsl{vs.} coupler frequency $\omega\ind{c}$ (black) and values from a numerical eigenvalue analysis based on the electrical parameters of the device (orange).
(b) Combined level- and pulse-diagram used for the measurement of the conditional phase. The control pulse on the coupler is a flat-top Gaussian, parametrized by the total length $\tau$ and the maximal amplitude $\Phi\ind{c}$.
}
\label{fig2}
\end{figure}

First, we characterize $\alpha\ind{\ZZ} (\omega\ind{c})$ as a function of coupler frequency (see Fig.~\ref{fig2}) by applying a static external magnetic flux $\Phi\ind{c}$, and by measuring the frequency of Q$_1$ in a Ramsey-type experiment for both cases of Q$_2$ being in the ground and the excited state.
The resulting frequency difference corresponds to $\alpha\ind{\ZZ}$ and spans over approximately 3 orders of magnitude $\alpha\ind{\ZZ} / 2 \pi = -0.06 \ldots 80 \unit{MHz}$ for the explored range of $\omega\ind{c}$ values. We find excellent agreement with the results from a numerical eigenvalue analysis based on the electrical circuit parameters of our device~\cite{supplemental}. 

We implement a two-qubit conditional phase gate by applying a flux pulse $\Phi\ind{c}(t)$ to tune the coupler frequency $\omega\ind{c}$. For simplicity, we use Gaussian-filtered square pulses, which are parametrized by their maximal amplitude $\Phi\ind{c}$, total pulse length $\tau$ (including zero-amplitude buffer elements) and the standard deviation $\sigma = 2.23 \unit{ns}$ of the Gaussian filter (see Fig.~\ref{fig2}(b) for a sketch of the pulse and Supplemental Material~\cite{supplemental} for an analytical expression of the pulse shape). The pulse is pre-distorted using a set of infinite impulse response (IIR) filters to account for the frequency-dependent transfer function of the flux drive line~\cite{supplemental}.

\begin{figure}[t!] 
\centering
\includegraphics[width = 0.48\textwidth]{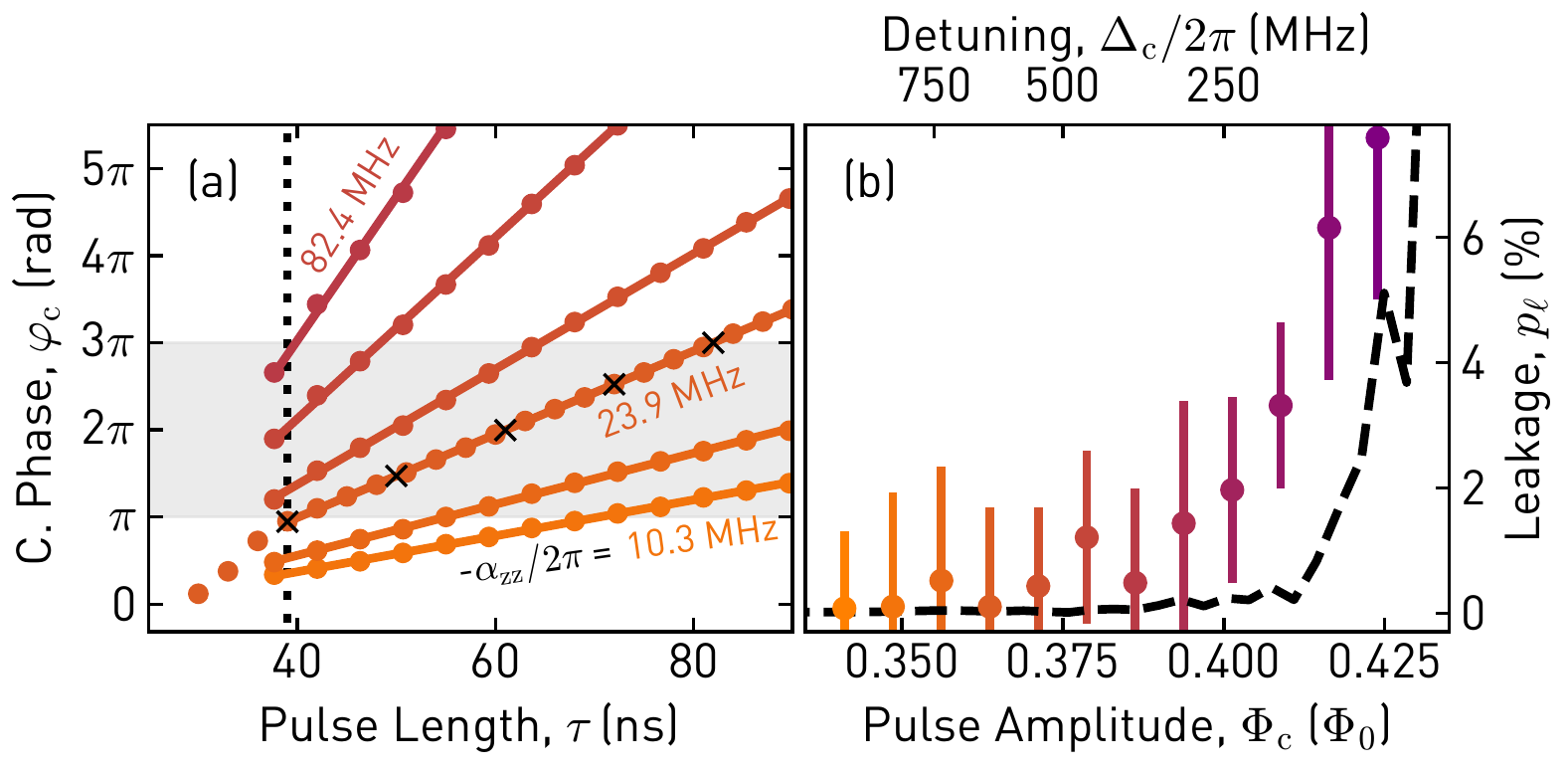}
\caption{(a) Measured conditional phase $\varphi\ind{c}$ \textsl{vs.} pulse length $\tau$ for various pulse amplitudes $\Phi\ind{c}$ (color coded, see b) and fit to linear model. Black crosses indicate the parameters for which we have performed quantum process tomography.
(b) Measured (colored points) and simulated (dashed line) accumulated leakage $p_{\ell}$ \textsl{vs.} pulse amplitude $\Phi\ind{c}$.
}
\label{fig3}
\end{figure}

Using a Ramsey-type experiment (see pulse scheme in Fig.~\ref{fig2}(b)), we measure the accumulated conditional phase $\varphi\ind{c}= \int \alpha\ind{zz}(\Phi\ind{c}(t))\, \mathrm{d}t$ acquired by Q$_2$ as a function of pulse length $\tau$ for different pulse amplitudes $\Phi\ind{c}$ (Fig.~\ref{fig3}(a)). We note a linear increase of $\varphi\ind{c} $ with $\tau$ for $\tau > 38\unit{ns}$.
The linear scaling is a direct consequence of the chosen square pulse shape and allows us to easily chose any targeted conditional phase by adjusting the pulse length accordingly.

To asses the quality of the conditional phase gate we perform quantum process tomography for the target phases $\varphi\ind{c}\dni{target} = \{\pi, \frac{3}{2} \pi, 2\pi, \frac{5}{2} \pi, 3\pi \}$ covering a range of $2\pi$ (gray highlight in Fig.~\ref{fig3}(a)). We find an average gate fidelity of $\mathcal{F}\ind{qpt} = 98.4\,\%$, with the best value $\mathcal{F}\ind{qpt}^{\pi} = 98.9\,\%$ observed for the shortest gate with $\tau = 38 \unit{ns}$ at a target phase of $\varphi\ind{c}\dni{target} = \pi$. The average required gate length is $\tau = 60 \unit{ns}$ for the chosen pulse amplitude of $\Phi\ind{c} = 0.37\, \Phi_0$. Shorter gates would be possible by relaxing the constraint to work in a regime with a simple linear dependence between $\varphi\ind{c}$ and $\tau$.

Next, we characterize the leakage properties of the coupling mechanism. To this aim, we determine the state of the full system $\ket{n_1, n_2, n\ind{coupler}}$, represented in the Fock basis, using simultaneous frequency multiplexed single shot readout~\cite{supplemental,walter_rapid_2017, heinsoo_rapid_2018}, and measure
the accumulated leakage $p_{\ell}$ into the non-computational states $\ket{200}$, $\ket{020}$,
$\ket{011}$,
$\ket{101}$,
and $\ket{001}$
as a function of pulse amplitude, see Fig.~\ref{fig3}(b).
For each pulse amplitude we perform 14 measurements with pulse lengths ranging from $38\unit{ns}$ to $94\unit{ns}$ and plot the mean value (dots) and the standard deviation (vertical lines).

For large pulse amplitudes $\Phi\ind{c} > 0.37 \, \Phi_0$ we find a substantial leakage, which we explain by the vanishing detuning $\Delta\ind{c} = \omega\ind{c} - \omega\ind{2}$ between the coupler and the qubit Q$_2$, leading to a non-negligible transverse coupling.
More refined pulse shapes assuring fast adiabatic control~\cite{martinis_fast_2014, li_realisation_2019, wang_experimental_2019} are expected to alleviate the effect of this  coupling and thus enable even faster gates.
For small amplitudes, however, we find leakage populations close to zero, within the systematic measurement uncertainty, which we estimate to be about $\Delta_{p} \approx 2\,\%$.
This trend is also observed in a master equation simulation (dashed line).

\begin{figure}[t!] 
\centering
\includegraphics[width = 0.48\textwidth]{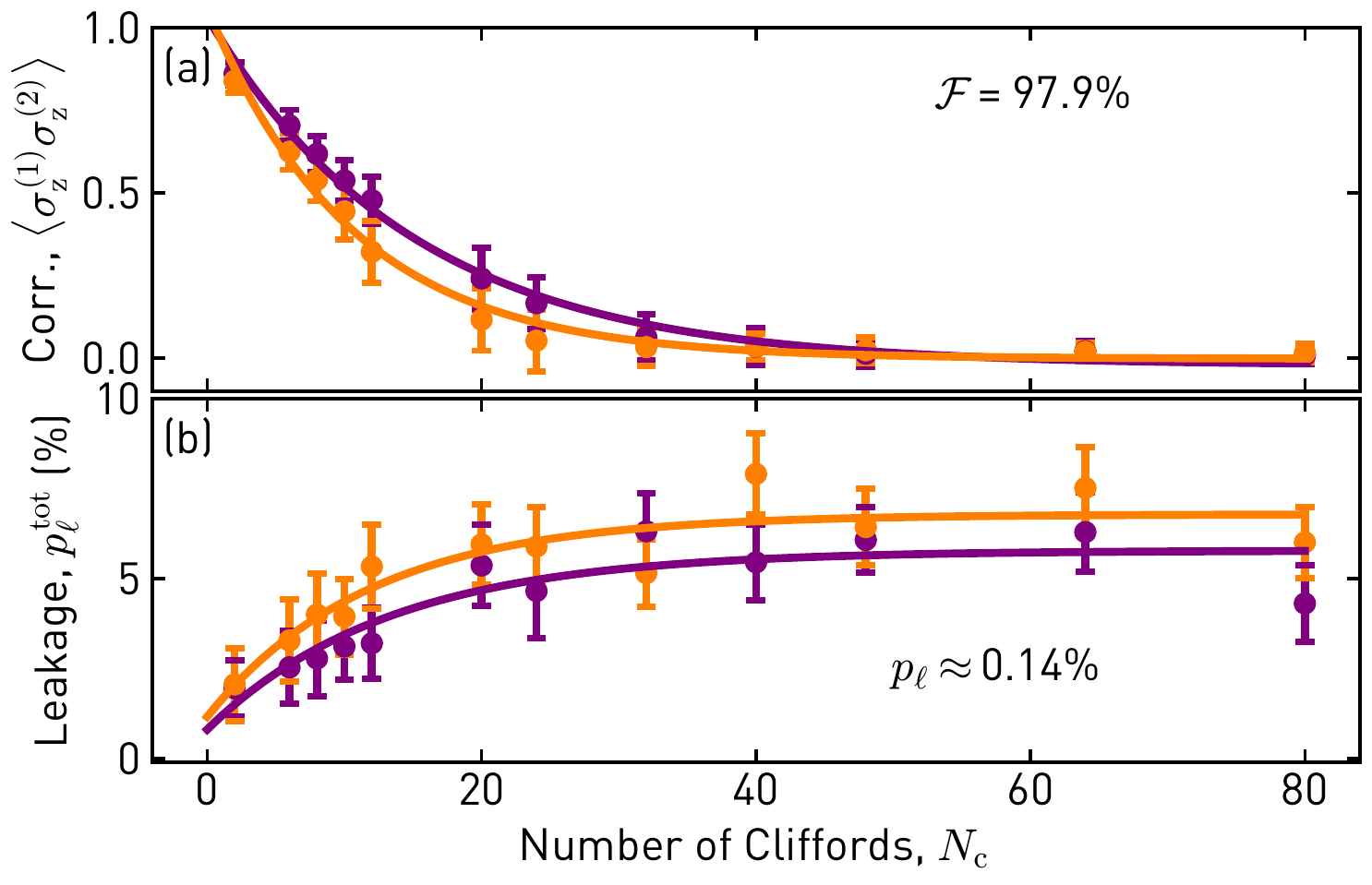}
\caption{
(a) Measured $\langle  \sigma\dni{(1)}\ind{z} \sigma\dni{(2)}\ind{z} \rangle$ and (b) accumulated leakage population $p_\ell\dni{tot}$ in randomized benchmarking (purple) and interleaved randomized benchmarking (orange) \textsl{vs.} sequence length $N\ind{c}$.
}
\label{fig4}
\end{figure}

In order to obtain a more precise estimate for the leakage per \CZ -gate in this regime we measure the leakage accumulated after applying multiple gates in (interleaved) randomized benchmarking sequences~\cite{magesan_efficient_2012,corcoles_process_2013} (see Fig.~\ref{fig4}(a)). We measure the $\langle \sigma\dni{(1)}\ind{z} \sigma\dni{(2)}\ind{z}\rangle$ correlator as well as the accumulated total leakage $p_{\ell}\dni{total}$ as a function of the number of elements chosen randomly from the two qubit Clifford group and extract a depolarization parameter per Clifford of  $r\ind{irb} = 0.91 \pm 0.01$ ($r\ind{rb} = 0.94 \pm 0.01$) for  sequences with (without) an additionally interleaved \CZ-gate per Clifford. This allows us to extract a fidelity of $\mathcal{F}\ind{\CZ} = (97.9 \pm 0.7)\, \%$
and an averaged leakage probability  of $p_{\ell} = (0.14 \pm 0.24) \, \%$ per \CZ -gate~\cite{chen_measuring_2016, wood_quantification_2018, rol_fast_2019}, see  Fig.~\ref{fig4}(b).

For comparison, we measure a fidelity $\mathcal{F}\ind{idle} > 99\,\%$ when interleaving a zero-amplitude flux pulse of length $\tau\dni{idle} = 40 \unit{ns}$ (see Supplemental Material~\cite{supplemental}), from which we conclude that the implemented \CZ -gate is not yet fully limited by decoherence.
We attribute this in part to coherent errors caused by imperfections in the calibration of the IIR filters and the resulting drift of the coupler frequency during the course of repeated gate applications.
This issue could be mitigated by improving the matching of the flux line or by using net-zero pulse shapes \cite{rol_fast_2019}. 

In conclusion, we have demonstrated a direct \ZZ -coupling between superconducting qubits, which is widely tunable over three orders of magnitude and thus well suited to implement conditional phase gates with high contrast.
This is expected to prove beneficial for mitigating the build-up of correlated errors between multiple qubits~\cite{krinner_benchmarking_2020, gutierrez_errors_2016}.
The ability to perform conditional phase gates for any target phase by simply tuning a single control parameter could be useful for substantially reducing the circuit depth in variational quantum algorithms~\cite{lacroix_improving_2020, abrams_implementation_2019, foxen_demonstrating_2020, ganzhorn_gate-efficient_2019}.
Moreover, in the prospect of engineered many-body systems of light~\cite{carusotto_quantum_2013, georgescu_quantum_2014, hartmann_quantum_2016, noh_quantum_2017}, rapid and precise control over nonlinear \ZZ -interactions in combination with linear transverse interactions offer unique prospects for the study of extended Bose-Hubbard models~\cite{roushan_spectroscopic_2017, kounalakis_tuneable_2018, jin_photon_2013}.

\begin{acknowledgments}
We thank Liangyu Chen, Philipp Kurpiers, Mihai Gabureac and Bruno K\"ung for discussions and experimental support.
This  work  is  supported
by the EU Flagship on Quantum Technology H2020-FETFLAG2018-03 project 820363 ``OpenSuperQ'',
by the Swiss National Science Foundation (SNSF) through the project ``Quantum Photonics with Microwaves in Superconducting Circuits'',
by the Office of the Director of National Intelligence (ODNI), Intelligence Advanced Research Projects Activity (IARPA), via the U.S. Army Research Office grant W911NF-16-1-0071,
by the National Centre of Competence in Research ``Quantum Science and Technology'' (NCCR QSIT), a research instrument of the Swiss National Science Foundation (SNSF),
and by ETH Zurich.
The views and conclusions contained herein are those of the authors and should not be interpreted as necessarily representing the official policies or endorsements, either expressed or implied, of the ODNI, IARPA, or the U.S.Government.
\end{acknowledgments}

\putbib
\end{bibunit}

\begin{bibunit}[apsrev4-1]

\widetext
\clearpage

\setcounter{equation}{0}
\setcounter{figure}{0}
\setcounter{table}{0}
\setcounter{page}{1}
\setcounter{secnumdepth}{3}
\makeatletter
\renewcommand{\theequation}{S\arabic{equation}}
\renewcommand{\thefigure}{\text{S}\arabic{figure}}
\renewcommand{\thetable}{\text{S}\Roman{table}}
\renewcommand{\bibnumfmt}[1]{[S#1]}
\renewcommand{\citenumfont}[1]{S#1}
\makeatother

\onecolumngrid

\begin{center}
\textsl{\textbf{\large{Supplemental Material to}}}
\vspace{0.05cm}

\centering\large{\textbf{\puttitle}}
\vspace{0.4cm}

\normalsize{
{Michele C. Collodo},$^{1,*}$
{Johannes Herrmann},$^{1,*}$
{Nathan Lacroix},$^1$
{Christian Kraglund Andersen},$^1$
{Ants~Remm},$^1$
{Stefania Lazar},$^1$
{Jean-Claude Besse},$^1$
{Theo Walter},$^1$
{Andreas Wallraff},$^1$ and
{Christopher Eichler}$^{1}$
}
\vspace{0.2cm}

\textit{\normalsize{
$^1${Department of Physics, ETH Zurich, CH-8093 Zurich, Switzerland}
}}
\vspace{1cm}
\end{center}

\twocolumngrid

\begin{figure*}[t] 
\centering
\includegraphics[width=0.96\textwidth]{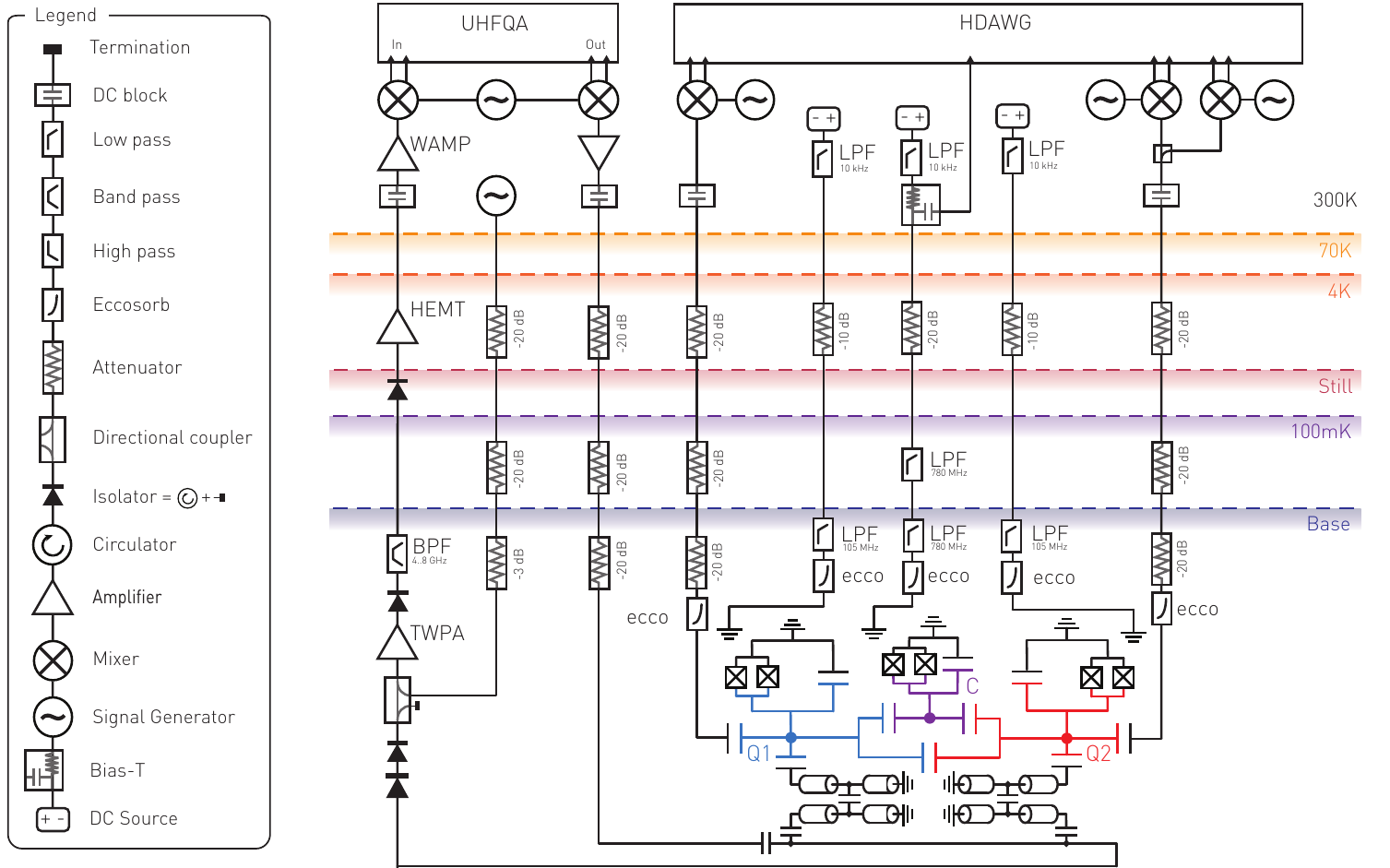}
\caption{Full wiring diagram of the experimental setup.}
\label{fig:setup}
\end{figure*}

\section{Experimental Setup}
\subsection{Wiring and Instrumentation}

We fabricated the sample using a Niobium thin film sputtered on Silicon in a process identical to the one described in Ref.~\cite{andersen_repeated_2019}.

We operated the sample at the base temperature ($20\unit{mK}$) of a cryogenic microwave setup and connected it to control and measurement electronics as shown in Fig.~\ref{fig:setup}.

To control the frequency of the qubits, flux drives applied to the SQUID loops of all three transmons are fed by a dc voltage source. Additionally, the fluxline to the coupler element C is controlled using a pulsed base-band signal (provided by a \textsl{Zurich Instruments HDAWG}), which is added to the dc signal by means of a bias-tee with a time constant on the order of $\tau\ind{bias-tee} \sim 18 \unit{\mu s}$.
We achive $XY$-control of the qubits by upconverting the intermediate frequency in-phase and quadrature AWG signals, with $\omega\ind{IF} /2\pi = 90\unit{MHz}$, to the respective microwave transition frequency using analog IQ-mixers.
For characterization purposes we apply pulses to the coupler by using the charge line of Q$_2$.
All drive pulses are generated from a single AWG featuring 8-channels with a sample rate of 2.4~GSa/s (\textsl{Zurich Instruments HDAWG}).

The baseband readout tone (see Section~\ref{readout} for a detailed description of the choice and parametrization of timing and frequency components) is generated and recorded by an FPGA-based control system with a sampling rate 1.8~GSa/s (\textsl{Zurich Instruments UHFQA}). This tone is up-converted to the target frequency of the respective readout circuit and amplified. The reflection off of the Purcell-filter-dressed readout resonator is routed to a near-quantum-limited traveling wave parametric amplifier (TWPA) \cite{macklin_nearquantum-limited_2015}. A bandpass filter restricts the signal to a $4...8 \unit{GHz}$ band before it is further amplified by a cryogenic high-electron-electron mobility transistor (HEMT) and microwave amplifiers at room temperature.
The signal is finally down-converted to an intermediate frequency, digitized and integrated using the weighted integration units of the \textsl{UHFQA}.

\subsection{Device Parameters}

\begin{table}[b]
\begin{tabular}{l c c c}
\hline
  & Q$_1$ & Q$_2$ &C\\ \hline \hline
Qubit frequency, $\omega_\mathrm{q}/2\pi$ (GHz)  &5.038&5.400&7.612\\
Lifetime, $T_1$($\mu$s) & 13.9 & 9.5 & 7.3 \\
Ramsey decoherence time, $T_2^*$($\mu$s) & 4.2 & 5.9 & 0.8 \\
Echo decoherence time, $T_2^\mathrm{echo}$($\mu$s) & 10.9 & 12.5 & 4.0 \\
Readout resonator frequency, $\omega_\mathrm{r}$(GHz) & 5.999 & 6.494 & - \\
Readout linewidth, $\kappa_\mathrm{eff}$(MHz) & 1.8 & 2.3 & - \\
Dispersive Shift, $\chi/2\pi$(MHz) & -2.5 & -2.6 & -0.7\footnote[1]{Dispersive sift on the readout resonator of qubit Q$_2$.} \\
Thermal population, $P_\mathrm{th} (\%)$ & 1.7 & 1.9 & 9.7 \\
 \hline
\end{tabular}
\caption{\label{tab:qb_paras} Measured qubit and readout parameters.}
\end{table}

We extract the qubit and readout parameters using standard spectroscopy and time domain measurements, summarized in Table~\ref{tab:qb_paras}.
Due to a noticeable thermal population, we condition the results of the time domain measurements on detecting all three elements in the ground state initially (see Section~\ref{readout} for a detailed description of the preselection method).

\section{Control and Readout Signals}
\subsection{Flux Pulse Parametrization}

The implementation of the two-qubit conditional phase gate requires precise frequency control of the coupler element, provided by a voltage pulse which is converted linearly to a magnetic flux pulse via the coupler fluxline.
To keep the total pulse length short and simplify the tuneup procedure we implement a flat-top Gaussian pulse shape
\begin{equation*}
\Phi_c(t) = \frac{\Phi_c}{2}\left(\mathrm{erf}\left(\frac{t-\tau\ind{b}}{\sqrt{2}\sigma}\right)-\mathrm{erf}\left(\frac{t-\tau\ind{c}-\tau\ind{b}}{\sqrt{2}\sigma}\right)\right),
\end{equation*}
with pulse amplitude $\Phi_c$, Gaussian filter width $\sigma=2.23\unit{ns}$, core pulse length $\tau\ind{c}$, and zero-amplitude buffer length $\tau\ind{b}$ to mitigate the influence of small mismatches in the timing of $Z$ and $XY$-pulses.
For the pulse with total length $\tau = 38 \unit{ns}$ presented in the main text, we choose $\tau\ind{c} = 14 \unit{ns}$ and $\tau\ind{b} = 12 \unit{ns}$. The latter is kept fixed for the pulse length sweeps.

To assure the repeatability of the pulse and obtain a flat frequency response of the coupler, we predistort the flux pulse waveform using infinite-impulse-response (IIR) filters.
We measure the instantaneous coupler frequency response to a step function and invert the obtained impulse response to calculate the respective IIR filter coefficients.
The filter tuneup procedure of our pulse requires the calibration of eight iteratively applied IIR filters with time constants ranging from $10\unit{\mu s}$ to $10~\unit{ns}$. To test the quality of the filter tuneup, we measure the conditional phase of $21$ subsequently repeated flux pulses and observe an average phase error of less than 1 degree per gate.

\subsection{Multiplexed Single Shot Readout}
\label{readout}

\begin{figure}[t] 
\centering
\includegraphics[width = 0.48\textwidth]{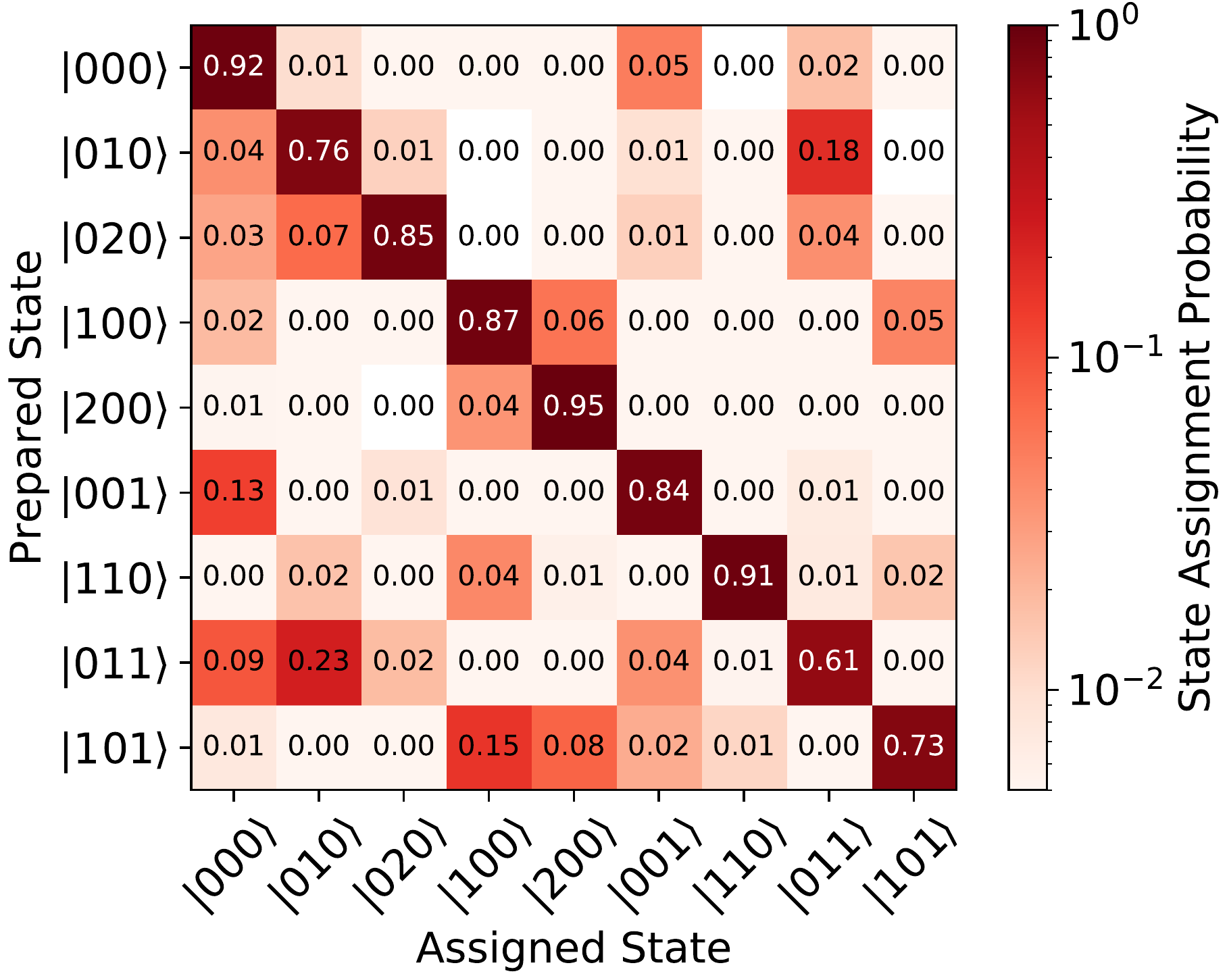}
\caption{State assignment probability matrix, showing the probability to assign the state indicated at the bottom axis when preparing the state on the left axis. All states are represented in the Fock-basis $\ket{n_1,n_2,n_\mathrm{coupler}}$. The state preparation includes a pre-selection readout to suppress preparation errors arising from thermal population.}
\label{fig:mro}
\end{figure}

Our single shot readout scheme \cite{walter_rapid_2017, heinsoo_rapid_2018} involves the classification of nine selected states from the one- and two-excitation manifold of the full qubit-coupler system.
The states included in the classification (see Fig.~\ref{fig:mro}) are chosen such that they optimally cover the most prominent leakage channels of the implemented conditional phase gate, which we identified from numerical master equation simulations using QuTip. These include the two-photon excitation on each qubit individually as well as the two-photon states comprised of one excitation on C and one excitation on Q$_1$ or Q$_2$, respectively.

Notably, readout of the coupler state does not require a dedicated third readout circuit due to a sufficient dispersive shift of a coupler excitation on both qubit readout resonators, see Table~\ref{tab:qb_paras}.
To be able to readout all elements in a frequency multiplexed fashion, we stimulate the readout circuitry with a pulse containing four frequency components $\omega_r/2\pi= (6.0132, 6.017, 6.4946, 6.5009)\unit{GHz}$ with amplitudes $v_r=(2.1, 2.2, 2.1, 2.2)\unit{mV}$ at room temperature, respectively.
The readout pulse has a total length of $250 \unit{ns}$ with a flat-top Gaussian envelope of $\sigma\ind{r}=25\unit{ns}$.

We prepare a set of all nine reference states and monitor the average time domain response of both in-phase and quadrature components of the reflected readout signal.

The eight non-zero differences between the responses of each prepared state and the ground state constitute the time-dependent integration weights specific to each individual state. In turn, this allows us to span an 8-dimensional phase space with the aforementioned set of integration weights as a basis. Additionally, we orthonormalize this basis by means of a Gram-Schmidt decomposition.
We then integrate the resonator response of each individual single shot readout-event for $600\unit{ns}$ using this set of eight pre-calculated optimal integration weights and collect it as a point in this space.
Subsequently, we train a Gaussian mixture model using the distribution of recorded reference responses, allowing us to assign each readout event to one of the nine predetermined states of the qubit-coupler system. The achieved assignment fidelity matrix $M$ is shown in Fig.~\ref{fig:mro}. We find the lowest fidelities for the states comprising one qubit and one coupler excitation, $\ket{011}$ and $\ket{101}$, due to both the reduced readout contrast of the coupler as well as the strong cross-Kerr nonlinearity on the order of $10\unit{MHz}$ between the coupler and the qubits. We prepend a further readout pulse before each experimental repetition and condition each run on having initially detected the ground state $\ket{000}$ with a probability larger than $99\,\%$ in order to minimize the effective thermal population.

Averaging over many realizations of an experiment results in a set of average excitation probabilities for each state $(p\ind{000}, p\ind{010}, ..., p\ind{110})$. In order to mitigate systematic imperfections of our readout scheme such as overlap error, we choose to correct the assigned average populations with the inverse of the assignment fidelity matrix, 
$(p\ind{000}\dni{cor}, p\ind{010}\dni{cor}, ..., p\ind{110}\dni{cor}) = M^{-1} \cdot (p\ind{000}, p\ind{010}, ..., p\ind{110})$.
This basis transformation relies predominately on the stability of the state assignment. In our experiments, the finite stability and number of repetitions of the state classification limits the accuracy of the reported probabilities to about $2\,\%$.

\section{Gate Characterization}
\subsection{Single Qubit Gate Performance}

\begin{figure}[t] 
\centering
\includegraphics[width = 0.48\textwidth]{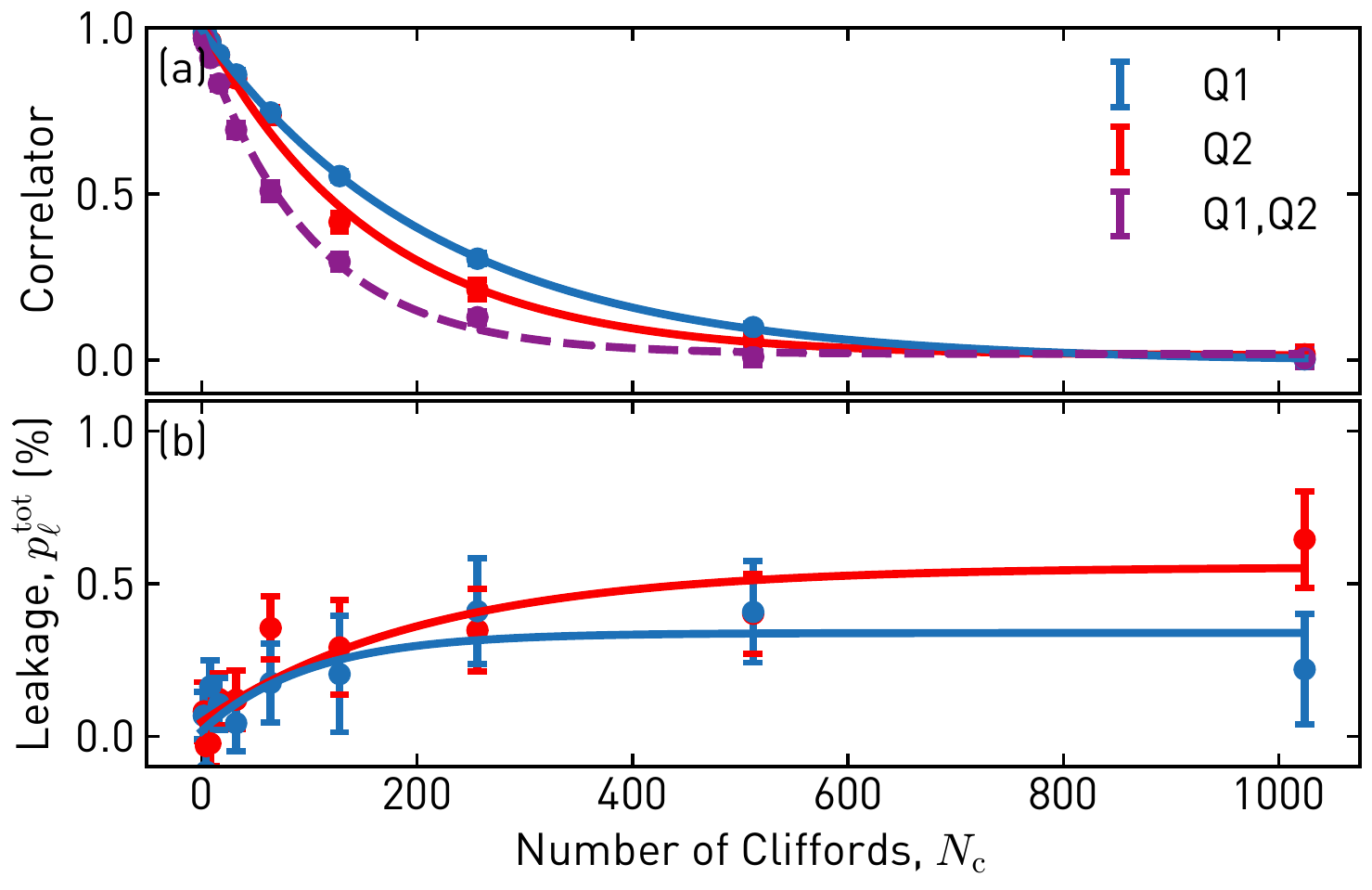}
\caption{Single qubit randomized benchmarking, with (a) measured correlators $\langle  \sigma\ind{z}\dni{(1)} \rangle$ (blue), $\langle  \sigma\ind{z}\dni{(2)} \rangle$ (red), and  $\langle  \sigma\ind{z}\dni{(1)} \sigma\ind{z}\dni{(2)} \rangle$ (purple) and (b) accumulated leakage $p_{\ell}\dni{tot}$ vs. sequence length~$N\ind{c}$. The red and blue data points are obtained from measuring the qubits individually, purple from a simultaneous measurement.}
\label{fig:srb}
\end{figure}

We characterize the single qubit gate performance by randomized benchmarking \cite{chen_measuring_2016} and find fidelities for each individual qubit of $(\mathcal{F}_{Q1},\mathcal{F}_{Q2})$ = (99.87, 99.83)\%, see Fig.~\ref{fig:srb}(a).
To assess the influence of control crosstalk, we perform a simultaneous single qubit RB experiment \cite{mckay_correlated_2020}, in which both individual gate sequences are applied to each qubit at the same time, and measure the correlator $\langle  \sigma\ind{z}\dni{(1)} \sigma\ind{z}\dni{(2)} \rangle$, equivalent to the two-qubit gate case. We extract a fidelity of $\mathcal{F}\ind{Q1Q2} = 99.73 \%$.

Additionally, we measure the residual population in the second excited states $\ket{020}$ and $\ket{200}$, respectively, after a randomized benchmarking sequence of length $N\ind{c}$ (see Fig.~\ref{fig:srb}b). From this we extract single qubit leakage rates per Clifford $(r_{Q1},r_{Q2})$=(3.4, 2.7)$\times 10^{-5}$. These values are approximately two orders of magnitude lower than the two-qubit gate leakage rates, ensuring that our results are not limited by single qubit control errors. 

\subsection{Comparison to Idle Gate}

\begin{figure}[t] 
\centering
\includegraphics[width = 0.48\textwidth]{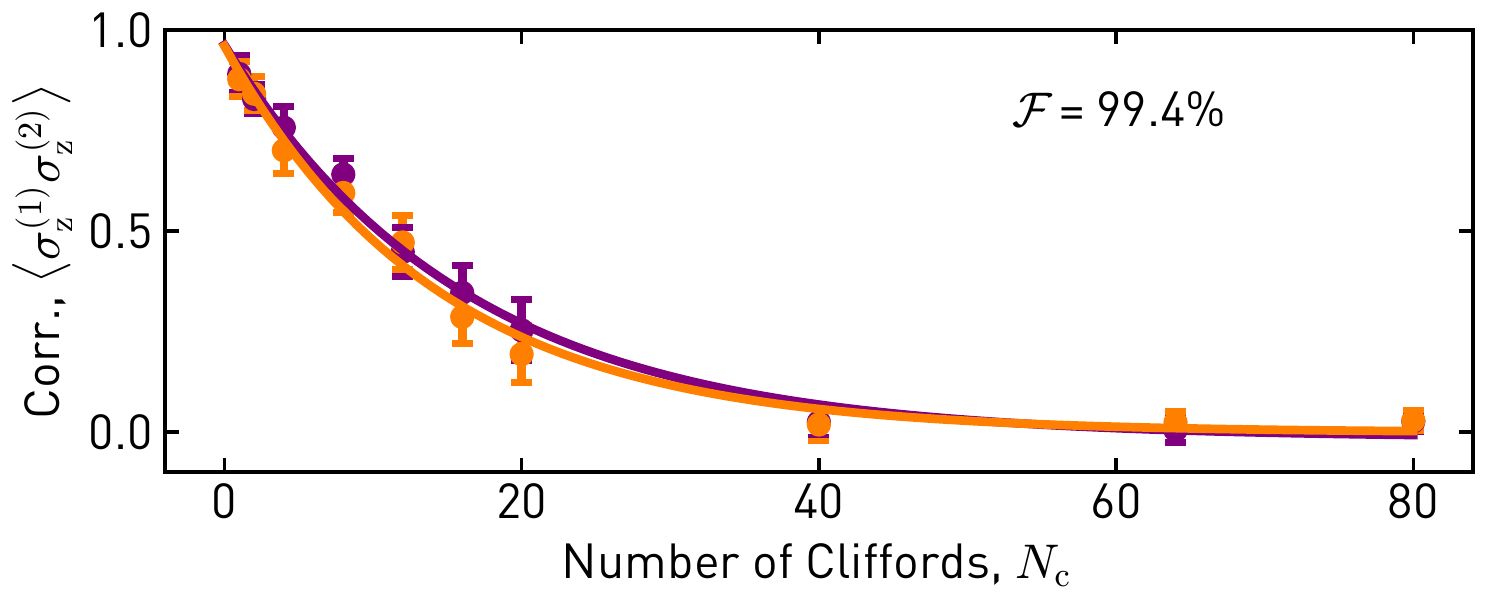}
\caption{Measured $\langle  \sigma\ind{z}\dni{(1)} \sigma\ind{z}\dni{(2)} \rangle$ in randomized benchmarking (purple) and interleaved randomized benchmarking (orange) \textsl{vs.} sequence length $N\ind{c}$. The interleaved Clifford element is an identity gate with a duration of $40 \unit{ns}$.}
\label{fig:idle}
\end{figure}

In order to assess the influence of the qubit decoherence rates on the final conditional phase gate, we perform a randomized benchmarking experiment and interleave the sequence with an identity operation with a duration of $40 \unit{ns}$ (see Fig~\ref{fig:idle}). Comparing the decay rates of the two qubit correlators $\langle  \sigma\ind{z}\dni{(1)} \sigma\ind{z}\dni{(2)} \rangle$ as a function of sequence length for the RB and iRB case yields a fidelity of $\mathcal{F}\ind{idle} = 99.39 \%$.
We thus conclude that the conditional phase gate presented in the main text is not fully coherence limited.
Coherent errors are likely accumulated due to imperfections in the calibration of IIR filters and the resulting skew of the coupler's operation frequency. This issue could be mitigated by improving the matching of the flux line or by using net-zero pulse shapes \cite{rol_fast_2019}. 

\begin{figure}[t] 
\centering
\includegraphics[width = 0.45\textwidth]{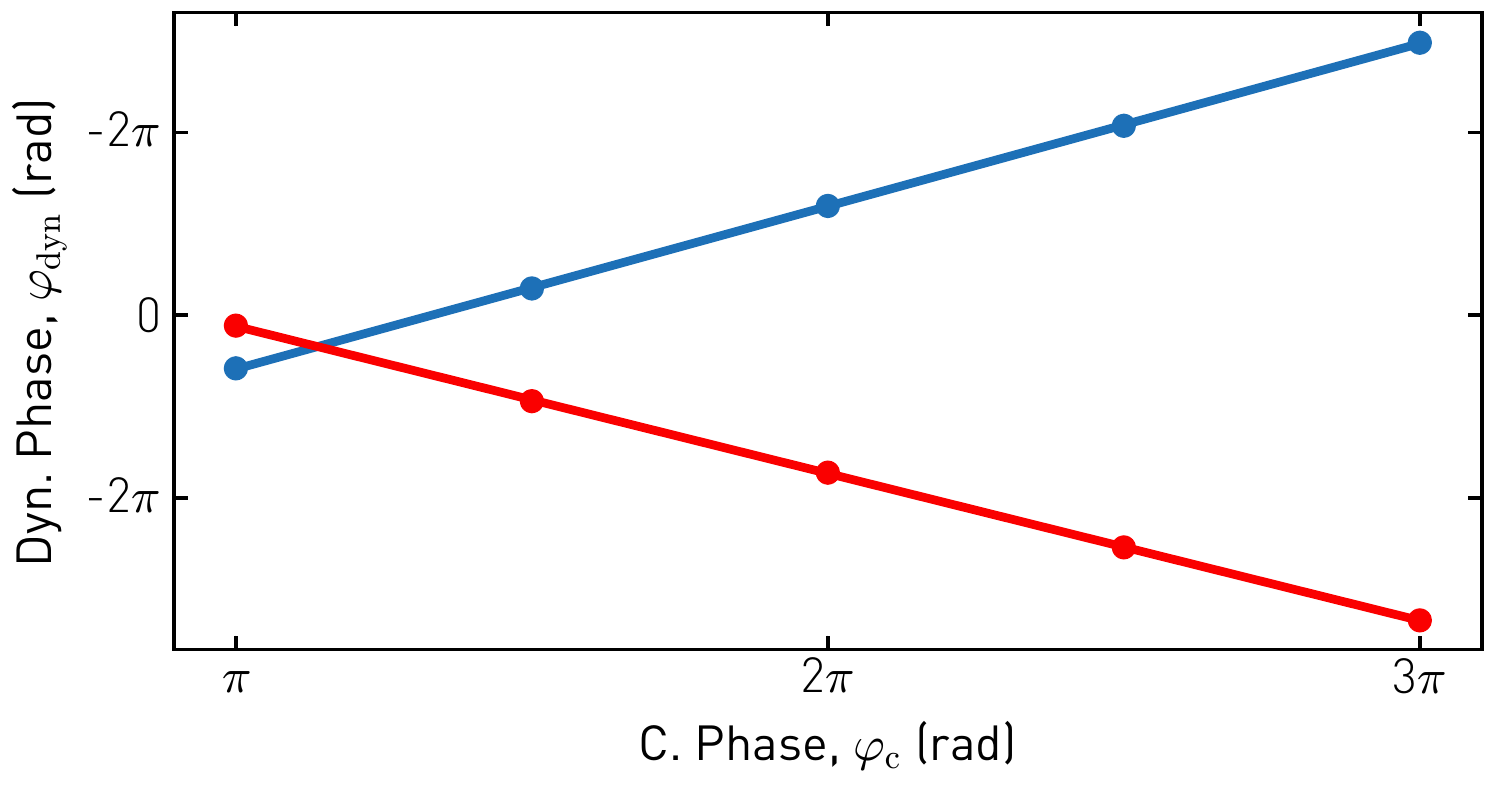}
\caption{Dynamical phase of Q$_1$ (blue) and Q$_2$ (red) vs. conditional phase. The measured data points are represented by colored dots and are overlayed with a linear fit.}
\label{fig:dyn_phase}
\end{figure}

\subsection{Quantum Process Tomography (QPT)}

Quantum Process Tomography allows us to assess the performance of a conditional phase gate with arbitrary target phase $\varphi\ind{c} \neq \pi$. We conducted QPT for five different phase angles $\varphi\ind{c}\dni{target} = (\pi, 3\pi/2, 2\pi, 5\pi/2, 3\pi)$ and find fidelities of $\mathcal{F_\mathrm{qpt}} = (98.9, 98.4, 97.6, 98.4, 98.6)\, \%$, respectively.
Here, we use the aforementioned readout mechanism and discard events exhibiting leakage out of the computational subspace. We use a maximum-likelihood estimation method to ensure the physicality of the reconstructed process matrix.

The operation of the gate relies on the frequency tuning of a coupler element.
However, strong capacitive couplings and finite flux crosstalk results in a non-negligible frequency excursion of the computational qubits. These frequency excursions lead to the accumulation of dynamic phases of the individual qubits, which we compensate with virtual single qubit \textsl{Z}-rotations.
The dynamic phase angles are generally dependent on the targeted conditional phase which we calibrate carefully.
We find that the acquired dynamic phase is linear in the conditional phase, see Fig~\ref{fig:dyn_phase}.
This is due to the robust variable phase gate implementation relying solely on the gate duration as a control parameter.

\section{Full Circuit Analysis}

\begin{figure}[t] 
\centering
\includegraphics[width = 0.35\textwidth]{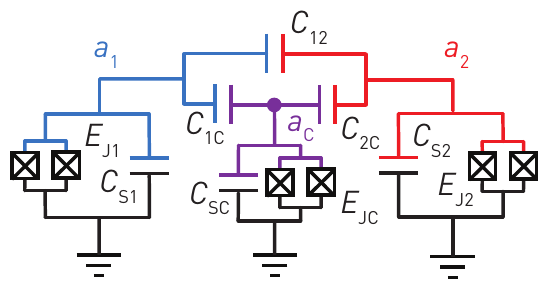}
\caption{Equivalent circuit diagram of the device shown in Fig.~\ref{fig1}, with Q$_1$ (blue), Q$_2$ (red), the coupler (purple) and the corresponding capacitive coupling network.}
\label{fig:circ}
\end{figure}

\begin{table}
\begin{tabular}{l c}
\hline
\hline
$C\ind{s1}$,$C\ind{s1}$  & $77.8 \unit{fF}$ \\
$C\ind{sc}$ & $60.4 \unit{fF}$\\
$C\ind{12}$ & $0.46 \unit{fF}$ \\
$C\ind{1c}$, $C\ind{2c}$ & $6.4 \unit{fF}$ \\
$E\ind{J1}$ & $h$ $15.3 \unit{GHz}$ \\
$E\ind{J2}$ & $h$ $17.49 \unit{GHz}$ \\
$E\ind{Jc}$ & $h$ $37.3 \unit{GHz}$ \\
$r$         & 1/1.71\\
\hline
\hline
\end{tabular}
\caption{\label{tab:electrical} List of electrical parameters.}
\end{table}

We model our device using the circuit diagram shown in Fig.~\ref{fig:circ}. The effective electrical parameters, encompassing corrections due to coupling capacitances to drive lines and readout circuitry, are listed in Tab.~\ref{tab:electrical}.

The circuit analysis is based on the Lagrangian
\begin{align*}
    \mathcal{L} &= \frac{1}{2} \dot \phi \, \mathcal{C} \, \dot \phi + \sum_{i =1,2} E_{\text{J}i} \cos \left(\phi_i / \Phi_0 \right) \\
    &+  \frac{E\ind{Jc}}{r+1} \sqrt{1+r^2 + 2 r \cos\left(\Phi\ind{c}/\Phi_0\right)} \  \cos \left(\phi\ind{c} / \Phi_0 \right)
\label{eq:legendre}
\end{align*}
with the phase coordiantes $\phi = (\phi_1, \phi_2, \phi\ind{c})$, the reduced flux quantum $\Phi_0$, and the capacitance matrix of the system
\begin{equation*}
\mathcal{C} = 
\begin{pmatrix}
C\ind{s1} + C\ind{12} + C\ind{1c} & -C\ind{12} & -C\ind{1c}\\
-C\ind{12} & C\ind{s2} + C\ind{12} + C\ind{2c} &  -C\ind{2c}\\
-C\ind{1c} & -C\ind{2c} & C\ind{sc} + C\ind{1c} + C\ind{2c}
\end{pmatrix}.
\end{equation*}
     
We attain the Hamiltonian in local modes by means of a Legendre transformation and subsequent Taylor expansion of the Cosine potential up to 6th order. We write the Hamiltonian in the Fock basis ($n=8$) and diagonalize it numerically.

The cross-Kerr coupling rate between the computational qubits (see Fig.~\ref{fig2} in the main text) is thus calculated as
$
    \alpha\ind{zz}(\omega\ind{c}) = \omega_{\ket{110}} - \omega_{\ket{100}} - \omega_{\ket{010}}
$,
where the $\omega_i$ are measured relative to the ground state energy.
Our calculation also takes fluxline crosstalk into account.

\begin{figure}[t] 
\centering
\includegraphics[width = 0.48\textwidth]{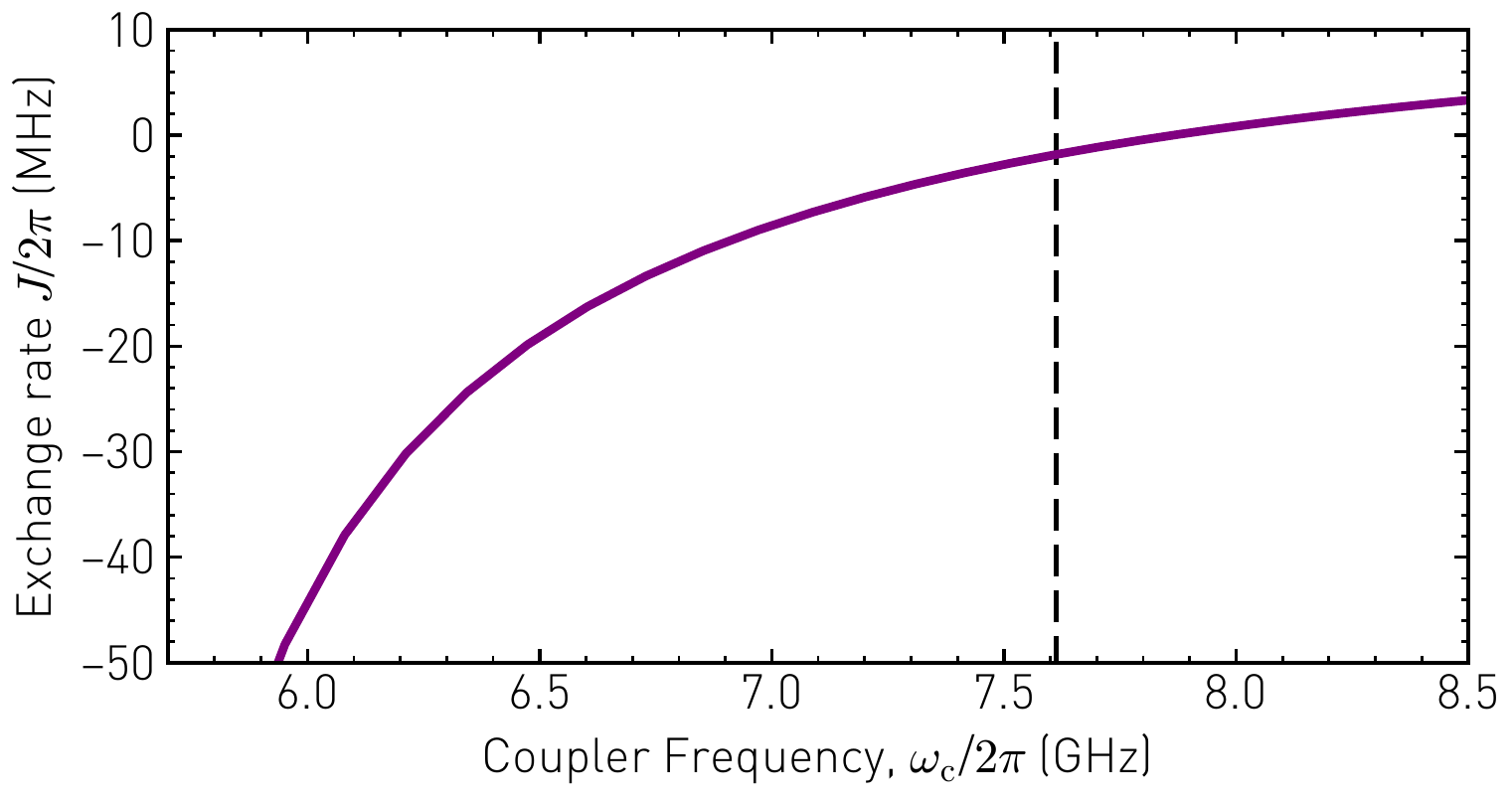}
\caption{Effective exchange interaction $J$ vs. coupler frequency $\omega\ind{c}$. The idle position of the coupler is indicated by a dashed line.}
\label{fig:J}
\end{figure}

To calculate the effective transverse coupling rate $J$ between the computational qubits (see Fig.~\ref{fig:J}), we choose $\tilde E\ind{J1} = \tilde E\ind{J2} = (E\ind{J1} + E\ind{J2})/2$ to attain identical frequencies for both qubit local modes. We then extract the $J$ rate as half the energy difference between the
 $\ket{100}$ and $\ket{010}$ eigenstates.
The transverse coupling between the computational qubits arises from the interference of the direct and the coupler-mediated coupling path and its magnitude and sign can be tuned by controlling the coupler frequency $\omega\ind{c}$ \cite{yan_tunable_2018}.

We simulate leakage to non-computational levels by time-evolving the system Hamiltonian using the Schr\"{o}dinger-equation solver of the QuTip library~\cite{johansson_qutip:_2012}. To increase simulation efficiency we apply the rotating-wave approximation to our Hamiltonian and thus can decrease the number of Fock states to $n=3$. 

\putbib
\end{bibunit}

\end{document}